\documentclass{iopart}

\usepackage{iopams} 
\usepackage{epsfig}
\usepackage{graphicx}
\usepackage{subfig}
\expandafter\let\csname equation*\endcsname\relax
\expandafter\let\csname endequation*\endcsname\relax 
\usepackage{amsmath}

\begin{document}

\title{Gluon Polarisation Measurements @ COMPASS}

\author{Lu\'{i}s Silva, on behalf of the COMPASS Collaboration}

\address{LIP -- Laborat\'{o}rio de Instrumenta\c{c}\~{a}o F\'{i}sica
  Experimental de Part\'{i}culas,\\Av. Elias Garcia, 14 1$^\circ$
  1000-149 Lisboa, Portugal}

\ead{lsilva@lip.pt}
\begin{abstract}
One of the missing keys in the present understanding of the spin
structure of the nucleon is the contribution from the gluons: the
so-called gluon polarisation. This quantity can be determined in DIS
through the photon-gluon fusion process, in which two analysis methods
may be used: (i) identifying open charm events or (ii) selecting
events with high $p_T$ hadrons. The data used in the present work were
collected in the COMPASS experiment, where a 160 GeV$/c$ naturally
polarised muon beam, impinging on a polarised nucleon fixed target is
used. Preliminary results for the gluon polarisation from high $p_T$
and open charm analyses are presented. The gluon polarisation result
for high $p_T$ hadrons is divided, for the first time, into three
statistically independent measurements at LO. The result from open
charm analysis is obtained at LO and NLO. In both analyses a new
weighted method based on a neural network approach is used.
\end{abstract}


\section{The Nucleon Spin}
\label{sec:Int}

The nucleon spin sum rule can be written in a heuristic way as:
$\frac{1}{2}=\frac{1}{2}\Delta \Sigma + \Delta G + L_{\mathit{q, g}}$,
where $\Delta \Sigma$ and $\Delta G$ are the quark and gluon
contributions to the nucleon spin, respectively, and $L_{\mathit{q,
g}}$ is the parton orbital angular momentum. In the late 80's it was
announced and confirmed from several experiments that the contribution
carried by the quarks is $\sim 1/3$ of the nucleon spin
\cite{emc,e1432,e142,smc,hermes}.  The purpose of this work is to
estimate the gluon polarisation $\Delta G/G$, which is deeply related
with the gluon contribution to the nucleon spin.

\section{Gluon Polarisation Measurements}
\begin{figure}[h!]
\begin{center}
\includegraphics[clip,width=0.8\textwidth]{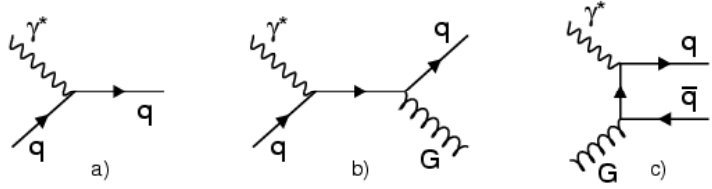}
\vspace{-10pt}
\caption{DIS Feynman diagrams for $\gamma^\star N$ scattering: a)
  virtual photo-absorption (LP), b) gluon radiation (QCD Compton) and
  c) photon-gluon fusion (PGF).}
\label{fig:figure1}
\vspace{-15pt}
\end{center}
\end{figure}

The spin dependent effects are measured experimentally using the
helicity asymmetry $A^{\rm exp}_{\rm LL}$ defined as
$\frac{\sigma^{\overleftarrow{\Leftarrow}}-\sigma^{\overleftarrow{\Rightarrow}}}{\sigma^{\overleftarrow{\Leftarrow}}+\sigma^{\overleftarrow{\Rightarrow}}}$
where ($\overleftarrow{\Leftarrow}$) and
($\overleftarrow{\Rightarrow}$) refer to the parallel and
anti-parallel spin helicity configuration of the beam lepton
($\leftarrow$) with respect to the target nucleon ($\Leftarrow$ or
$\Rightarrow$). The data for the present analyses was taken in the
\textsc{Compass} experiment \cite{compass}. The gluon polarisation can
be measured via the Photon-Gluon Fusion (PGF) process, depicted in
\fref{fig:figure1} c), which allows to probe the spin of the gluon
inside the nucleon. The PGF process may be selected using two analysis
methods: (i) selecting high $p_T$ hadron events, or (ii) selecting
events containing open charm mesons.

\subsection{High $p_T$}
\label{sec:hiptHI}
In the high $p_T$ analysis the spin helicity asymmetry is calculated
by selecting events containing high $p_T$ hadron pairs above $0.7$ and
$0.4$ GeV$/c$, respectively for the highest and the second highest
$p_T$ hadron with respect the virtual photon direction. A cut on $Q^2
> 1 ~(\mathrm{GeV}/c)^2$ was also applied in order to select DIS. Two
other processes compete with the PGF process in LO QCD approximation,
namely the virtual photo-absorption leading order process (LP) and the
gluon radiation (QCD Compton) process, illustrated in
\fref{fig:figure1}. The spin helicity asymmetry for the high $p_T$
hadron pair data sample can thus be schematically written as:
\begin{equation}
A_{\rm LL}^{2h}(x_{Bj})= R_{\rm PGF} \, a_{\rm LL}^{\rm
PGF}\frac{\Delta G}{G}(x_G) + R_{\rm LP} \, D \, A_1^{\rm LO}(x_{Bj})
+ R_{\rm QCDC} \, a_{\rm LL}^{\rm QCDC} A_1^{\rm LO}(x_C)~.
\label{eq:allmain}
\end{equation}
The process fractions are represented by $R_i$, $i$ referring to the
different processes. $a_{\rm LL}^i$ represents the partonic cross
section asymmetries, $\Delta\hat{\sigma}^i/\hat{\sigma}^i$, also known
as analysing power. The depolarisation factor $D$ is the
fraction of the muon beam polarisation transferred to the virtual
photon. The virtual photon asymmetry $A_1^{\rm LO}$ is defined as
$A_1^{\rm LO} \equiv \frac{\sum_i e_i^2 \Delta q_i}{\sum_i e_i^2
q_i}$. A similar equation to \eref{eq:allmain} can be
written to express the inclusive asymmetry of a data sample, $A_{\rm
LL}^{incl}$. Using \eref{eq:allmain} for the high $p_T$ hadron pair
sample and the above mentioned equation for the inclusive sample the
final expression to extract the gluon polarisation is obtained:
\begin{equation}
  \frac{\Delta G}{G}(x_G^{av}) = \frac{A_{\rm
      LL}^{2h}(x_{Bj})+A^\mathrm{corr}}{\lambda}~.
  \label{eq:dgg} 
\end{equation}
This formula corresponds to the spin helicity asymmetry $A_{\rm
LL}^{2h}$, measured directly from data, plus a correcting asymmetry
$A^\mathrm{corr}$ involving mainly the virtual photo-absorption and
the gluon radiation processes. The $\lambda$ factor relates the
partonic asymmetries and the fractions of the involving
processes. During the data selection process it is not possible to
identify the process that originated each event, neither to access its
partonic variables. Therefore the partonic asymmetries and the process
fractions need to be estimated using a dedicated and well tuned Monte
Carlo (MC) simulation. The quality of the simulation is illustrated by
the comparison of the distributions for three variables in the
presented at the conference slides.  In the high $p_T$ analysis a
Bayesian Neural Network (NN) \cite{Sulej:2007zz} approach was
used. The purpose of such approach is to assign a event probability
for each process involved: LP, QCD Compton and PGF. These
probabilities represent the process fractions in
\eref{eq:allmain}. Also the NN is used to provide the partonic
asymmetries and $x_C$ and $x_G$ variables. The gluon polarisation is
calculated in an event by event basis using an optimal weight which
improves the figure of merit. Details about the high $p_T$ analysis,
for $Q^2 > 1 ~(\mathrm{GeV}/c)^2$, can be found in \cite{hiptHQ}.  A
similar analysis was performed for the $Q^2 < 1 ~(\mathrm{GeV}/c)^2$
data, which contains $\sim 90\%$ of the whole $Q^2$ range. This
separation is due to the physical processes contained in the two $Q^2$
regimes. The $Q^2 < 1 ~(\mathrm{GeV}/c)^2$ regime represents the
quasi-real photon, in such conditions the photon may exhibit some
inner structure. Therefore beside the already mentioned three
intervening processes the inclusion of photon structure processes in
the MC simulation needs to be accomplished. Details of this analysis
can be found in \cite{Ageev:2005pq}.

\subsection{Open Charm}
\label{sec:charm}
For the open charm analysis the spin helicity asymmetries are
calculated using a data with $D^0$ mesons in the final state. These
events are selected from their decaying products, \textit{i.e.} $K\pi$
pairs. To achieve this selection a good particle identification is
required. Applying a set of kinematic cuts the combinatorial
background originated from process in which the virtual photon strikes
a parton inside the nucleon is reduced. In addition the backgroun is
suppressed by selecting the $D^\star \to D^0 \pi_\mathit{slow}$
channel. Doing in this way three channels were included in the final
analysis: $D^0 \to K \pi$, $D^0 \to K \pi \pi^0$ and $D^0 \to K \pi
\pi \pi$. The data used in this analysis are deuteron data collected
from 2002 to 2006 and proton data collected in 2007.

The number of events with $D^0$ particles in the final state is related with the
helicity asymmetries as shown by this expression:
\begin{equation}
  \label{eq:N}
N_t=\alpha (S+B) \left[ 1 + \beta \left( a_{\rm LL} \frac{S}{S+B}
    \frac{\Delta G}{G} + D \frac{B}{S+B} A^{\rm bg}\right)\right]~. 
\end{equation}
The subscript $t$ on the number of events corresponds to the possible
muon target spin configurations. The $\alpha$ factor contains the
acceptance, muon flux and number of nucleons and $\beta$ the beam and
target polarisations and dilution factor. $S$ and $B$ represent the
number of signal and background events taken under the invariant mass
spectrum peak. $\frac{A(B)}{S+B}$ is the signal (background)
significance. $A^{\rm bg}$ is the asymmetry of the background. Taking
into account all the possible muon target spin configurations and the
unknown variables a set of equations is derived from the expression
\eref{eq:N}. This equation set is solved by $\chi^2$ minimisation. As
in high $p_T$ analysis, the gluon polarisation calculation is
performed event by event with an appropriate weight. Still to solve
this equation system the partonic asymmetry $a_{\rm LL}$ and the
signal significance $\frac{S}{S+B}$ must be estimated for every
event. To compute the partonic asymmetry $a_{\rm LL}$ a dedicated MC
simulation is used. A NN approach is designed to parametrise the
partonic asymmetry and the signal significance $\frac{S}{S+B}$ as a
function of the kinematics. The latter represents the event
probability for having $D^0$ particles. Details about this analysis
can be found in \cite{Franco:2010zz}.  A NLO QCD analysis was also
performed. Into the analysing power NLO QCD virtual and gluon
bremstrahlung corrections were included to the PGF process, as well as
background processes.

\section{Results}
The preliminary results on gluon polarisation using the high $p_T$  ($Q^2 < 1$
 and $Q^2 > 1 ~(\mathrm{GeV}/c)^2$, LO QCD order) and open charm (LO
 and NLO QCD order) analyses are now presented.  The $\Delta G/G$
 value obtained in high $p_T$ analysis, for $Q^2 > 1
~(\mathrm{GeV}/c)^2$, averaged at $x_g=0.09^{+0.08}_{-0.04}$  was
found to be equal to
$\Delta G/G= 0.125\pm0.060_{stat}\pm0.063_{sys}$. This measurement is
presented in three statistically independent points, in \tref{tb:dgg}.
The same result for $Q^2 < 1 ~(\mathrm{GeV}/c)^2$ is $\Delta G/G=
0.016\pm0.058_{stat}\pm0.055_{sys}$. 

\begin{table}[h!]
\vspace{-15pt}
\caption{\label{tb:dgg}Gluon polarisation results in bins of $x_G$.}
\begin{center}
\begin{tabular}{llll}
\br &$1^{st}$ Bin&$2^{nd}$ Bin&$3^{rd}$ Bin\\ \mr $\Delta
G/G$&$0.147\pm 0.091\pm 0.088$&$0.079\pm 0.096\pm 0.082$&$0.185\pm
0.165\pm 0.143$\\
$x_G^{av}$&$0.07_{-0.03}^{+0.05}$&$0.10_{-0.04}^{+0.07}$&$0.17_{-0.06}^{+0.10}$\\
\br
\end{tabular}
\end{center}
\vspace{-15pt}
\end{table}

\begin{figure}[h!]
\vspace{-15pt}
\centerline{
\subfloat[LO QCD]{\label{fig:figure2a}\includegraphics[width=0.5\textwidth]{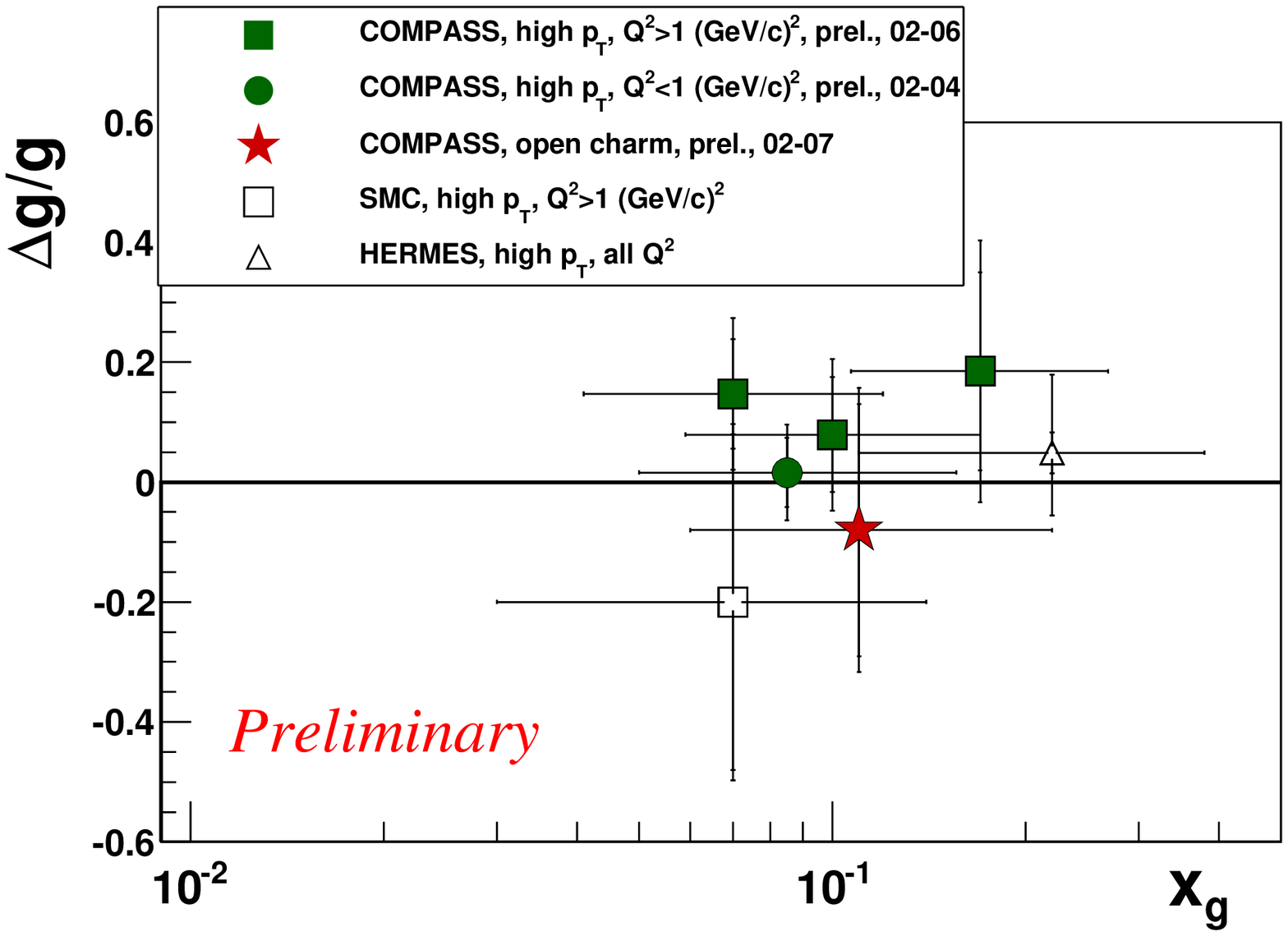}}
\subfloat[NLO QCD]{\label{fig:figure2b}\includegraphics[width=0.5\textwidth]{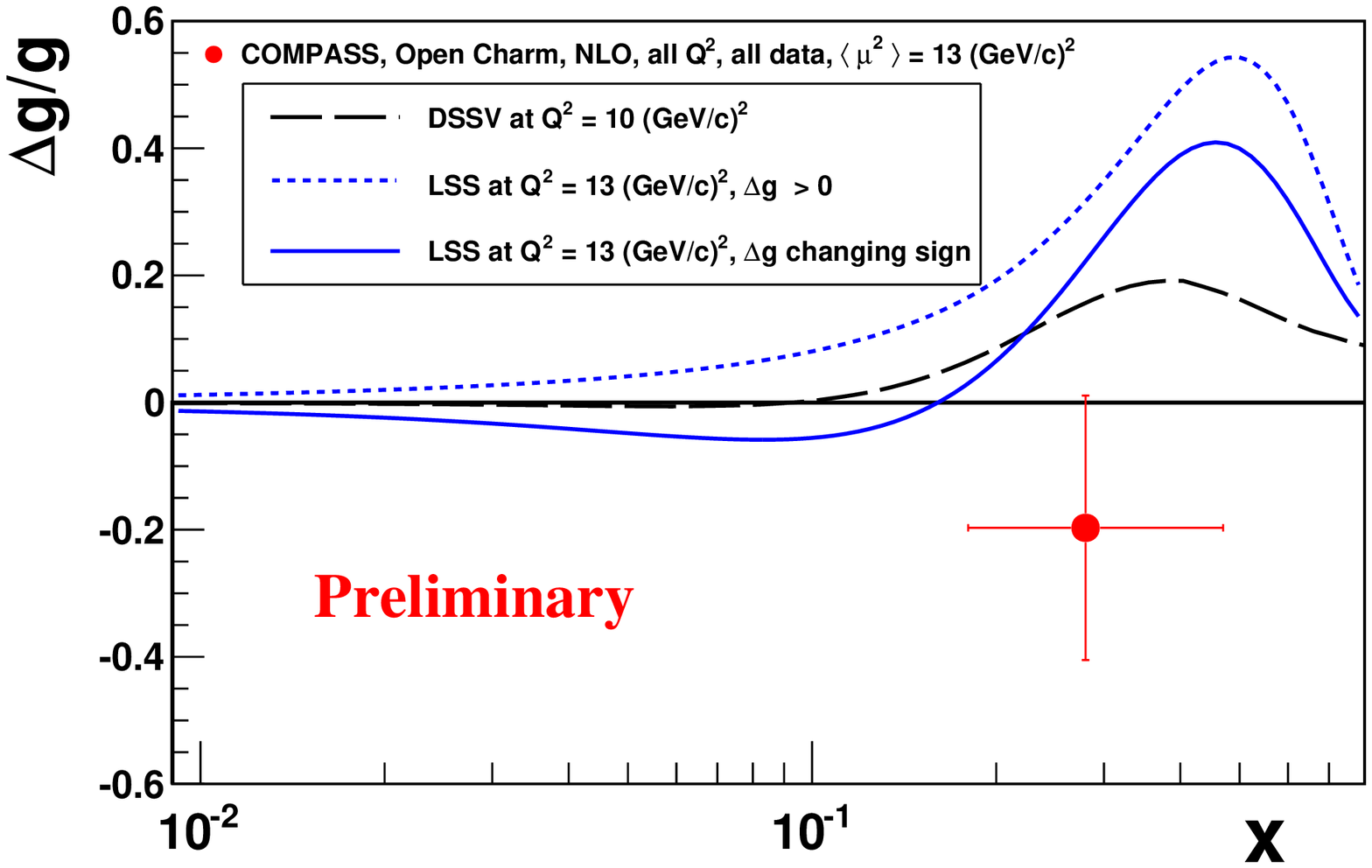}}}
\vspace{-10pt}
\caption{Gluon polarisation results: \subref{fig:figure2a} LO QCD
  order results,  \subref{fig:figure2b} NLO QCD order results.}
\vspace{-15pt}
\label{fig:figure2}
\end{figure}

The gluon polarisation value for the open charm analysis was found as
$\Delta G/G= -0.08\pm0.21_{stat}\pm0.08_{sys}$ averaged at
$x_G=0.11^{+0.11}_{-0.05}$ for LO QCD order and $\Delta G/G_{NLO}=
-0.20\pm0.21_{stat}\pm0.08_{sys}$ averaged at
$x_G=0.28^{+0.19}_{-0.10}$ at NLO QCD order. The gluon polarisation
values from high $p_T$ hadrons are measured at a hard scale of
$\langle\mu^2\rangle= 3.4 ~(\mathrm{GeV}/c)^2$, while the ones from
open charm mesons were evaluated at a hard scale of
$\langle\mu^2\rangle= 13 ~(\mathrm{GeV}/c)^2~.$ All the gluon
polarisation results of the \textsc{Compass} Collaboration are
summarised in \fref{fig:figure2} together with the SMC and HERMES
results.

More details can be found on the conference slides 

(\texttt{http://www.nuclear.kth.se/NCNP2011/Presentation\_files/L.Silva.pdf}).

\end{document}